\documentclass[
 aip,
 amsmath,amssymb,
preprint,%
]{revtex4-1}

\usepackage{graphicx}
\usepackage{dcolumn}
\usepackage{bm}

\usepackage[utf8]{inputenc}
\usepackage[T1]{fontenc}
\usepackage{mathptmx}
\usepackage{etoolbox}

\usepackage{xcolor}
\usepackage[normalem]{ulem} 

\makeatletter
\def\@email#1#2{%
 \endgroup
 \patchcmd{\titleblock@produce}
  {\frontmatter@RRAPformat}
  {\frontmatter@RRAPformat{\produce@RRAP{*#1\href{mailto:#2}{#2}}}\frontmatter@RRAPformat}
  {}{}
}%

\begin{document}

\preprint{}

\title{Kinetic versus Magnetic Chaos in Toroidal Plasmas: A systematic quantitative comparison}

\author{H.T. Moges}
 \affiliation{Nonlinear Dynamics and Chaos Group, Department of Mathematics and Applied Mathematics, University of Cape Town, Rondebosch 7701, South Africa}
 
\author{Y. Antonenas}
\affiliation{School of Applied Mathematical and Physical Sciences, National Technical University of Athens, Athens 15780, Greece}

\author{G. Anastassiou}
\affiliation{School of Applied Mathematical and Physical Sciences, National Technical University of Athens, Athens 15780, Greece}

\author{Ch. Skokos}
\affiliation{Nonlinear Dynamics and Chaos Group, Department of Mathematics and Applied Mathematics, University of Cape Town, Rondebosch 7701, South Africa}

\author{Y. Kominis}
\affiliation{School of Applied Mathematical and Physical Sciences, National Technical University of Athens, Athens 15780, Greece}
\email{gkomin@central.ntua.gr}
\date{\today}

\begin{abstract}
Magnetic field line chaos occurs under the presence of non-axisymmetric perturbations of an axisymmetric equilibrium and is manifested by the destruction of smooth flux surfaces formed by the field lines. These perturbations also render the particle motion, as described by the guiding center dynamics, non-integrable and, therefore, chaotic. However, the chaoticities of the magnetic field lines and the particle orbits significantly differ both in strength and radial location in a toroidal configuration, except for the case of very low-energy particles whose orbits closely follow the magnetic field lines. The chaoticity of more energetic particles, undergoing large drifts with respect to the magnetic field lines, crucially determines the confinement properties of a toroidal device but cannot be inferred from that of the underlying magnetic field. In this work, we implement the Smaller
ALignment Index (SALI) method for detecting and quantifying chaos, allowing for a systematic comparison between magnetic and kinetic chaos. The efficient quantification of chaos enables the assignment of a value characterizing the chaoticity of each orbit in the space of the three constants of the motion, namely energy, magnetic moment and toroidal momentum. The respective diagrams provide a unique overview of the different effects of a specific set of perturbations on the entire range of trapped and passing particles, as well as the radial location of the chaotic regions, offering a valuable tool for the study of particle energy and momentum transport and confinement properties of a toroidal fusion device.
\end{abstract}

\maketitle

\section{\label{sec:Inroduction} Introduction}

The configuration of the magnetic field is of crucial importance for the particle confinement in fusion plasmas. In tokamaks, the MagnetoHydroDynamic (MHD) equilibrium is designed to have an axisymmetric magnetic field \cite{FreidbergBook}- however, this background magnetic field is always perturbed by non-axisymmetric perturbations of various origins. Spontaneously induced instabilities result in stochastic transport of particle energy and momentum as well as particle loss, and can significantly compromise the performance of a fusion device \cite{WhiteBook, white2021particle}, whereas intentionally applied external perturbations in the form of Resonant Magnetic Perturbations (RMP) can be utilized for plasma control and suppression of instabilities \cite{shinohara2018estimation, xu2018loss, he2019resonant, shinohara2020efficient, he2020roles, he2020full}.  

A crucial question concerns the effect of each specific set of perturbations on the particle transport, in order to explore synergies and mitigation techniques under a perturbation engineering approach. A non-axisymmetric perturbation results in magnetic field line chaoticity and destruction of the well-defined concentric closed flux surfaces of the magnetic equilibrium \cite{rechester1978electron, biewer2003electron, WhiteBook, abdullaev2008description}. The specific flux surfaces that are mostly perturbed and destroyed are dictated by a resonance condition depending on the radial profile of safety factor $(q)$ and the poloidal $(m)$ and toroidal $(n)$ mode numbers of the perturbing non-axisymmetric mode having a generic spatial dependence of the form $a_{mn}\exp[i(m\theta-n\zeta)]$. 
Chains of magnetic islands are formed in the position of the resonant flux surfaces, with chaos being localized close to separatrices, for small perturbation strengths. Larger perturbation strengths result in spatially extended chaos, with the dynamics of the magnetic field lines being determined by the complex structure of the homoclinic and heteroclinic lobes of the unstable (saddle) periodic field lines \cite{Evans2005}.

Low-energy particles, with velocities mostly parallel to the magnetic field, closely follow the magnetic field lines, so that the chaoticity of their orbits and their transport characteristics are in a direct relation to the form of the magnetic field lines. In such cases, magnetic and kinetic chaos take place at the same spatial positions. However, higher-energy particles have considerable drifts across the magnetic field lines and their orbits may span different magnetic field lines. Such orbits may partially reside either inside or outside the regions where the magnetic field topology has been drastically changed with the formation of magnetic island chains or extended chaotic regions, but their dynamics are, in general, radically different from those of the magnetic field \cite{ram2010dynamics, cambon2014chaotic, ram2014anomalous, mitra2014particle, ogawa2016full, samanta2017energization}. For energetic particles, kinetic and magnetic chaos may occur at different positions, depending strongly on the kinetic characteristics of the particle (energy, parallel and perpendicular momentum). Moreover, the degree of chaoticity can significantly vary for different particles. These effects are of crucial importance to the dynamics of specific species of energetic particles such as energetic ions \cite{wang2022effects}, thermal \cite{morton2018electron} and runaway electrons \cite{carbajal2020runaway}. For example, the presence of magnetic perturbations has been shown to crucially impact the fast ion current profiles, by altering their kinetic characteristics as described by reduced “kick models” utilized for interpretive and predictive analysis within the framework of the integrated tokamak transport codes \cite{Podesta2019, Bardoczi2019}.  It is worth noting that, although minority particle species can be treated as passive tracers whose motion may not radically affect the structure of the magnetic field, the motion of thermal (bulk) electrons may alter the distribution of plasma current that, in turn, will affect the magnetic configuration. Similarly, in the advanced stages of a disruption, runaway electrons can carry most of the plasma current, so their orbits will influence the magnetic field structure \cite{Matsuyama2014, carbajal2020runaway}. In such cases, the full plasma dynamics can be described by a self-consistent treatment, which is beyond the scope of this work.

The Hamiltonian formalism provides the appropriate context and techniques for the study of magnetic and kinetic chaos; in fact, several advances in Hamiltonian methods have been motivated by these applications \cite{AbdullaevBook, LL92Book}. The zero divergence of the magnetic field implies the representation of the magnetic field lines as a one-degree-of-freedom Hamiltonian system, with the toroidal angle playing the role of time. The system is either autonomous or non-autonomous depending on the axisymmetry of the configuration. Particle motion can be studied in terms of its Guiding Center (GC) motion under the Hamiltonian formalism utilizing canonical variables defined by the background magnetic field \cite{littlejohn1983variational, WhiteBook}.

For axisymmetric magnetic fields, both Hamiltonian systems, describing the magnetic field lines and the GC motion, are integrable. However, the presence of axisymmetry-breaking perturbations results in the destruction of an invariant of each system and renders them both non-integrable. The effect of the perturbation on the phase space of the two systems is strongly inhomogeneous due to its resonant character, and therefore, chaos appears only in regions where  resonance conditions are fulfilled. Strong perturbations may result in resonance overlap and extended chaotic regions, but still these regions usually occupy only a part of the phase space. The localized character of phase space chaoticity gives rise to the question of the specific location of chaos. It is worth emphasizing that the phase space location does not refer only to a spatial location, such as the radial distance from the magnetic axis (related to the magnetic flux $\psi$ and canonical poloidal particle momentum $P_\theta$), but also to the values of the kinetic characteristics of the particles, namely their energy $E$, magnetic moment $\mu$, and canonical toroidal momentum $P_\zeta$, being Constants Of the Motion (COM) of the unperturbed GC Hamiltonian. The latter uniquely label each particle orbit and determine the specific characteristics of the particles that are actually affected by the presence of the perturbation, in terms of their categorization as trapped or passing, and confined or lost. This information is critical for studying the effect of each perturbation on the particle energy and momentum transport and confinement.

A visual identification of chaotic field lines and particle orbits can be performed by constructing Poincar{\'e} surfaces of section \cite{LL92Book}. Relatively low-energy particles tend to follow magnetic field lines so that kinetic and magnetic Poincar{\'e} surfaces of section are similar. However, for more energetic particles, there is no obvious relation between kinetic and magnetic chaos; in fact, kinetic chaos can be stronger in spatial locations where the magnetic chaos is weaker, and vice versa. The systematic investigation of the relation between kinetic and magnetic chaos necessitates the utilization of an index for efficient chaos detection and quantification.

The aim of this work is to systematically compare the chaoticity of magnetic field lines to the chaoticity of particle orbits due to non-axisymmetric magnetic perturbations. Chaos detection is based on the calculation of the  Smaller ALignment Index (SALI) \cite{skokos2001alignment}, utilized for the first time in the context of plasma physics. The calculation of such a quantitative measure of chaos enables the characterization of particle orbits in terms of their chaoticity in the COM space of the system and provides a compact overview of the effect of a specific perturbation on particles with different kinetic characteristics. 

The paper is organized as follows: The Hamiltonian models for the description of the magnetic field lines and the GC particle motion are presented in Section \ref{sec:Hamiltonian models}. Chaos detection and quantification in terms of the SALI method is introduced in Section \ref{sec:Chaos Detection, Quantification and Measure} for the Hamiltonian systems of interest and its advantages with respect to other chaos indices (such as the maximum Lyapunov Exponent - mLE) are discussed. Magnetic and kinetic chaos are systematically quantified and compared in Section \ref{sec:Magnetic versus Kinetic Chaos in Toroidal Plasmas}, where the SALI is depicted in the COM space of the GC motion to clearly characterize the chaoticity of different particle orbits (trapped/passing, confined/lost). Finally, the main findings and the conclusions are summarized in Section \ref{sec:Summary and Conclusion}.

\section{Hamiltonian Models \label{sec:Hamiltonian models}}

\subsection{\label{sec:Hamiltonian description of the magnetic field lines}Hamiltonian description of the magnetic field lines}

A canonical Hamiltonian description for the lines of a magnetic field $\mathbf{B}$ with general toroidal topology can be given in terms of Boozer coordinates \cite{boozer1981plasma} as 
\begin{equation}\label{gen mag field} 
    \mathbf{B} = \nabla\psi\times\nabla\theta - \nabla\psi_{p}\times\nabla\zeta,
\end{equation}
where $\theta$ and $\zeta$ are the toroidal and poloidal angles, $\psi$ and $\psi_{p}$ are the toroidal and poloidal fluxes of the magnetic field. It can be shown that the equations for the magnetic field lines can be written in a Hamiltonian form as \cite{WhiteBook}
\begin{align}\label{unper Ham MFL}
    \frac{d\psi}{d\zeta} = -\frac{\partial \psi_{p}}{\partial\theta}, && \frac{d\theta}{d\zeta} = \frac{\partial \psi_{p}}{\partial\psi},
\end{align}
where the poloidal flux $\psi_{p}(\psi, \theta,\zeta)$ has the role of the Hamiltonian, the toroidal angle $\zeta$ has the role of time, $\theta$ is the canonical coordinate and $\psi$ is its conjugate canonical momentum. The Hamiltonian system has one degree of freedom and it is non-autonomous, and therefore in general non-integrable, when the Hamiltonian $\psi_p$ depends explicitly on the toroidal angle $\zeta$. Independence of $\psi_p$ on $\zeta$ corresponds to an axisymmetric magnetic field configuration with $d\psi_{p}/d\psi\equiv1/q(\psi)$, defining the safety factor $q(\psi)$. From Eq.~\eqref{unper Ham MFL} it is clear that $q(\psi)=d\zeta/d\theta$ defines the helicity of the magnetic field on a magnetic surface of a given $\psi$.  

An explicit dependence of the Hamiltonian $\psi_p$ on the poloidal and toroidal angles strongly modifies the topology of the magnetic field lines, due to the localized or extended destruction of the magnetic surfaces \cite{LL92Book, WhiteBook, AbdullaevBook}. In such cases, the field lines do no longer form smooth constant $\psi$ surfaces and can become chaotic. The explicit dependence on the angles results from the presence of perturbative terms that modify the Hamiltonian as   
\begin{equation} \label{eq:HAM-MFLInteg}
    \psi_p = \int \frac{d\psi}{q(\psi)} -\sum_{m,n}a_{mn}(\psi) \sin(m\theta- n\zeta),
\end{equation}
where $a_{mn}(\psi)$ describes the flux (radial) profile of the perturbative mode, and $m, n$ are the mode numbers. Thus, Eqs.~\eqref{unper Ham MFL} become
\begin{align}\label{per Ham MFL}
    \frac{d\psi}{d\zeta} = \sum_{m,n}ma_{mn}(\psi) \cos(m\theta-n\zeta), && \frac{d\theta}{d\zeta} = \frac{1}{q(\psi)}.
\end{align}
The integrable part of the Hamiltonian [Eq.~\eqref{eq:HAM-MFLInteg}] is defined solely by the safety factor profile, having the general analytic form
\begin{equation} \label{gen q factor}
    q(\psi) = q_{ma} \Bigg[1 + \bigg(\bigg(\frac{q_{w}}{q_{ma}}\bigg)^\nu - 1\bigg)(\psi/\psi_{w})^\nu \Bigg]^{1/\nu},
\end{equation}
where $q_{ma}$ and $q_w$ are the values of $q$ at the magnetic axis and the wall, respectively, $\psi_w$ is the corresponding toroidal flux (associated with the radial position) at the wall, and $\nu$ controls the radial shape of the $q$-profile \cite{WhiteBook}. The $q$-profile determines the local helicity of the unperturbed field lines on a constant $\psi$ magnetic surface. Depending on the relation of the local field line helicity with the helicity of a specific mode ($m/n$) a resonance condition can be met, resulting in the destruction of the specific magnetic surface, the formation of a magnetic island chain and the appearance of local chaoticity of the field lines. Extended chaotic regions may occur under conditions for neighboring magnetic island overlap, according to Chirikov's criterion \cite{LL92Book}, and may significantly modify plasma transport as well as the operational conditions and the overall performance of the fusion device. It is worth noting that extended chaoticity of the magnetic field lines is not a necessary condition for chaotic particle transport and losses, as has been shown, for example, in the case of 10 keV electrons under the presence of tearing modes in a pre-disruptive ITER scenario \cite{Spizzo2019, Spizzo2009}, and also discussed in this paper.

\subsection{\label{sec:Guiding Center Hamiltonian}Guiding Center Hamiltonian}

The Hamiltonian GC theory in canonical variables \cite{white1984hamiltonian, WhiteBook} provides a reduced dynamical description in which a charged particle's fast gyro-motion about a local magnetic field line is asymptotically separated from slower bounce and drift motions along and across magnetic field lines. The GC motion of a charged particle in a  magnetic field is described by the Littlejohn Lagrangian \cite{littlejohn1983variational},
$L = (\mathbf{A}+\rho_{||}\mathbf{B}) \cdot \mathbf{u}+\mu\dot{\xi}-H$, where $\mathbf{A}$ and $\mathbf{B}$ are the vector potential and the magnetic field, $\mathbf{u}$ is the GC velocity, $\mu$ is the magnetic moment, $\xi$ is the gyro-phase, $\rho_{||}$ is the velocity component parallel to the magnetic field, normalized to B, and 
\begin{equation}\label{GC H}
    H = \frac{\rho_{||}^2}{2} B^2 + \mu B,
\end{equation}
is the Hamiltonian corresponding to the particle energy $E$. The GC motion is given in normalized units where time is normalized to $\omega_{0}^{-1}$, with $\omega_{0}=q_iB_{0}/m_i$ the on-axis gyro-frequency, and distance is normalized to the major radius $R$, so that energy is normalized to $m_i\omega_{0}^2R^{2}$, where $q_i$ and $m_i$ are the ion charge and mass, respectively. For different particle species, the GC drift velocities scale with the mass-to-charge ratio. According to the ordering of the GC approximation \cite{WhiteBook}, the gyro-radius is $\rho=u/B<<1$, and the magnetic moment $\mu=u_{\bot}^2/(2B)$ as well as the cross-field drift are of order $\rho^2$.

In the limit, where perturbations are approximated by linear superposition on the background field B, perturbations whose displacement is mainly perpendicular to the unperturbed flux surfaces, such as tearing or [Alfv{\'e}n modes, can be described as \cite{White2013a, White2013b, WhiteBook}
\begin{equation}
    \delta \mathbf{B} = \nabla\times\alpha\mathbf{B},
\end{equation}
where $\alpha = \sum_{m,n} \alpha_{m,n}(\psi) e^{i(m\theta-n\zeta)}$ and the Lagrangian is expressed in magnetic field line (Boozer) coordinates as
\begin{equation} \label{Lan}
    L = \left[\psi + (\rho_{||} + \alpha) I\right]\Dot{\theta} +\left[(\rho_{||} + \alpha) g - \psi_{p}\right]\Dot{\zeta} + \mu\Dot{\xi} - H.
\end{equation}
This form of the Lagrangian, derived after the omission of a term related to the non-orthogonality of the coordinate system \cite{WhiteBook, Bierwage2022}, readily provides the three couples of canonical conjugate variables $(P_{\theta},\theta)$, $(P_{\zeta},\zeta)$ and $(\mu,\xi)$, with
\begin{align}\label{canon moments}
    P_{\theta} = \psi + \left(\rho_{||} + \alpha \right) I, &&  P_{\zeta} = \left(\rho_{||} + \alpha \right) g - \psi_{p},
\end{align}
and the Hamiltonian [Eq.~\eqref{GC H}] is expressed in canonical variables as
\begin{equation} \label{per H}
    H = \frac{\left[P_{\zeta} + \psi_{p}(\psi) - \alpha(\psi,\theta,\zeta)\right]^2}{2} B^2(\psi,\theta) + \mu B(\psi,\theta),
\end{equation}
where $\psi$ and $\psi_p$ are functions of the canonical momenta $P_\theta, P_\zeta$ defined by Eq.~\eqref{canon moments}.

In the following we consider a Large Aspect Ratio (LAR) cylindrical equilibrium \cite{FreidbergBook,WhiteBook} described, to leading order, by $g\simeq1$, $I\simeq 0$, and $B=1-\sqrt{2\psi}\cos{\theta}$, where the magnetic field amplitude is normalized to its on axis value. In this case, the canonical momentum $P_\theta$ is equal to the toroidal flux $P_\theta=\psi$, and the equations of motion are written as 
\begin{align}\label{dot coordinate}
      \dot{\theta} = \frac{1}{q(P_{\theta})}\rho_{||}B^2 + \frac{1}{q(P_{\theta})}(\mu+\rho_{||}^2B)\frac{\partial B}{\partial \psi_{p}}, && \dot{\zeta} = \rho_{||}B^2,
\end{align}
\begin{align}\label{dot momentum}
    \dot{P}_{\theta} = -(\mu+\rho_{||}^2 B)\frac{\partial \alpha}{\partial \theta} + \rho_{||}B^2\frac{\partial \alpha}{\partial \theta}, && \dot{P}_{\zeta} = \rho_{||}B^2\frac{\partial \alpha}{\partial \zeta}.
\end{align}
It can be readily shown that low-energy particles, with $\mu \rightarrow 0$ and $\rho_{||} \rightarrow 0$, have negligible drifts and actually follow the magnetic field lines so that  
the equations of motion correspond to the equations for the magnetic field lines [Eq.~\eqref{per Ham MFL}].

In the absence of perturbations $(\alpha=0)$ the GC Hamiltonian system is axisymmetric and therefore integrable with three independent COM, $(E, P_{\zeta}, \mu)$ labeling each unperturbed orbit. In the COM space, orbits can be classified with respect to being trapped or passing, and confined or lost \cite{WhiteBook}. For the case of a LAR configuration the loss boundary is given by 
\begin{equation}\label{walls}
    E = \frac{\left(P_{\zeta} + \psi_{p}(\psi_{w})\right)^2}{2}\left(1\mp\sqrt{2\psi_{w}}\right)^2 + \mu\left(1\mp\sqrt{2\psi_{w}}\right),  
\end{equation}
where signs $-$ and $+$ correspond to particles co-moving and counter-moving with respect to the magnetic field. Moreover, orbits which pass through the magnetic axis ($\psi=0$) satisfy the equation
\begin{equation}\label{magnetic axis}
    E = \frac{P_{\zeta}^2}{2} + \mu.  
\end{equation}
and the boundary separating trapped and passing particles is given by 
 \begin{equation}\label{trapped passing boundary}
     E = \mu\left(1\mp\sqrt{2\psi(P_{\zeta})}\right).
 \end{equation}
Eqs. \eqref{walls}-\eqref{trapped passing boundary} define parabolas in the $(E, P_{\zeta})$ plane and allow for the characterization of each orbit in a constant $\mu$ slice of the three-dimensional (3D) COM space. The corresponding diagrams provide a valuable tool for the systematic quantitative comparison of magnetic and kinetic chaos and display a clear view for the specific type of particles (trapped/passing, co-/counter-moving) that are affected under the presence of perturbations, and will be utilized in the following analysis. 

\section{\label{sec:Chaos Detection, Quantification and Measure}Chaos Detection, Quantification and Measure}

The chaotic nature of a system is determined by how originally nearby initial conditions diverge asymptotically from each other. Over the years several numerical techniques have been developed to study this behavior: Poincar{\'e} surfaces of section \cite{LL92Book} for visualizing the dynamics of systems with a low number of degrees of freedom; the so-called `converse KAM method' for the identification of KAM tori \cite{mackay1985converse, Kallinikos2023}; the computation of quantities related to the orientation of small orbital perturbations (see for example \cite{contopoulos1997fast}, and \cite{white2012modification, White2015, WhiteBook} for the so-called ‘phase vector rotation
method'); the computation of various quantities like the mLE \cite{benettin1980lyapunovA, benettin1980lyapunovB, skokos2009lyapunov}, the Fast Lyapunov Indicator (FLI) \cite{froeschle1997fasta, froeschle1997fastb}, as well as the Smaller (SALI) and the Generalized (GALI) Alignment Index \cite{skokos2001alignment,skokos2007geometrical,skokos2016smaller} for detecting chaos. Moreover, several techniques have been developed for efficiently computing these chaos indicators (e.g. tangent map method \cite{skokos2010numerical}). The interested reader is referred to \cite{skokos2016chaos} where review papers of some of these methods are included and to a recent work where the performance of several chaos detection techniques is investigated \cite{bazzani2023performance}. 

The chaotic behavior of orbits in nonlinear dynamical systems has long been quantified by the estimation of the mLE ($\Sigma$) which practically measures the average divergence or convergence rate of the distance between two neighboring orbits in a dynamical system's phase space, and it is usually estimated by \cite{skokos2009lyapunov} 

\begin{equation}\label{eq:mLE}
\Sigma = \lim\limits_{t\to \infty} \sigma(t), 
\end{equation}
where $\sigma(t)$ is the so-called finite-time mLE (ftmLE) given by

\begin{equation}\label{eq:ftmLE}
\sigma(t) = \frac{1}{t}\ln \frac{\lVert{\mathbf{w}(t)}\Vert}{\lVert{\mathbf{w}(0)}\Vert}.
\end{equation}
$\mathbf{w}(0)$ and $\mathbf{w}(t)$ are deviation vectors from a given orbit, at times $t=0$ and $t>0$, respectively, and $\lVert{.}\Vert$ denotes the usual Euclidean norm of a vector. For the Hamiltonian system the mLE is greater than zero if the motion is chaotic, while it is equal to zero if the motion is regular \cite{skokos2009lyapunov}. In particular, in the latter case, the ftmLE tends to zero following a power law decay $\sigma(t) \propto t^{-1}$.

The estimation of the mLE has been successfully used in plasma physics for many years as a chaos indicator, as well as for the detection of Lagrangian Coherent Structures related to transport barriers \cite{falessi2015lagrangian, veranda2017magnetohydrodynamics, Giannatale2018a, Giannatale2018b, veranda2019helically}. However, the slow convergence of the ftmLE to its limiting value, caused by the fact that $\sigma(t)$ depends on the whole evolution of $\mathbf{w}(t)$ remains a disadvantage, notably for weakly chaotic orbits. For this reason, we chose to implement the SALI method in this work since it has been proven to be very efficient in detecting chaos faster than the estimation of the mLE, and it has been effectively used, for more than two decades, in investigations of several dynamical systems (see more on references  \cite{skokos2016smaller,skokos2010numerical, manos2008application, chaves2017boxy}), but has never been used in the study of magnetic and kinetic chaos in toroidal plasmas.

To compute the SALI of an orbit one should follow the time evolution of the orbit itself and of two initially linearly independent deviation vectors $\mathbf{w_1}(t)$ and $\mathbf{w_2}(t)$, having for example perpendicular directions \cite{skokos2001alignment, skokos2016smaller}. The two deviation vectors $\mathbf{w_1}(t)$ and $\mathbf{w_2}(t)$ are normalized at each time, t, such that 
\begin{equation}\label{2 dev vect}
\mathbf{\hat{w}_i}(t) = \frac{\mathbf{w_i}(t)}{\lVert \mathbf{w_i}(t) \Vert}, \qquad i =1,2.
\end{equation}

Then, the SALI is computed at time $t$ by \cite{skokos2001alignment} 
\begin{equation}
\mbox{SALI}(t) = min \Big \{ \lVert \mathbf{\hat{w}_1}(t) + \mathbf{\hat{w}_2}(t) \Vert, \lVert \mathbf{\hat{w}_1}(t) - \mathbf{\hat{w}_2}(t) \Vert \Big \}.
\label{eq:SALI}
\end{equation}
We note that SALI $=0$ indicates that the two deviation vectors, $\mathbf{w_1}(t)$ and $\mathbf{w_2}(t)$ are aligned, while the index takes its maximum value,  SALI$=\sqrt{2}$, when the two vectors are perpendicular. In the case of regular orbits, the two deviation vectors tend to fall on a torus' tangent space pointing at distinct directions, resulting in SALI having a positive, practically constant value. For chaotic orbits, however, the two deviation vectors align in the direction defined by the mLE, and SALI tends exponentially fast to zero at the rate associated to the two largest Lyapunov exponents, $\sigma_1$ and $\sigma_2$ defined by the relation \cite{skokos2004detecting}
	\begin{equation} 	\label{Prop:SALI_C}
	\mbox{SALI}(t) \propto e^{-(\sigma_1 - \sigma_2)t}.
	\end{equation}

For regular orbits in the case of 2D area-preserving maps, the two deviation vectors tend to fall on the 1D torus' tangent space, which is also 1D. Thus, the two vectors will eventually become linearly dependent, resulting in SALI asymptotically tending to zero, as in the case of chaotic orbits, but at a different time rate given by \cite{manos2007studying,skokos2016smaller} 
\begin{equation} \label{Prop:SALI-SM}
\mbox{SALI}(t) \propto n^{-2},
\end{equation}
where $n$ is the number of iterations of the 2D map.

In our study, we numerically solved the equations of motion for the Hamiltonians of the GC [Eqs.~\eqref{dot coordinate} - Eq.~\eqref{dot momentum}] and magnetic field lines [Eq.~\eqref{per Ham MFL}], along with the corresponding variational equations by using the fourth-order Runge-Kutta integrator, also employed in widely used orbit-following codes like ORBIT \cite{white1984hamiltonian} and ASCOT \cite{Hirvijoki2014}, with appropriately chosen integration steps, which kept the absolute relative energy errors less than $10^{-8}$.

In order to explain in detail how the chaos indicators considered in our study identify chaos, we consider, for simplicity, a radially uniform (independent of $\psi$) additive perturbation term in the LAR guiding center Hamiltonian and a constant safety factor, $q(\psi)=1$. Hence, the GC Hamiltonian is given by: 

\begin{equation}
H = \frac{(P_\zeta + \psi_p(P_\theta))^2}{2} B^2 + \mu B + \epsilon \left[ \sin (m_1\theta - n_1 \zeta) + \sin (m_2\theta - n_2\zeta) \right],
\label{eq:GCM-Ham-q=1}
\end{equation}
with $B=1-\sqrt{2P_\theta} \cos \theta$ and $\psi_{p}(P_{\theta}) = P_{\theta}$.

\begin{figure}
	\centering
	\includegraphics[width=1\textwidth,keepaspectratio]{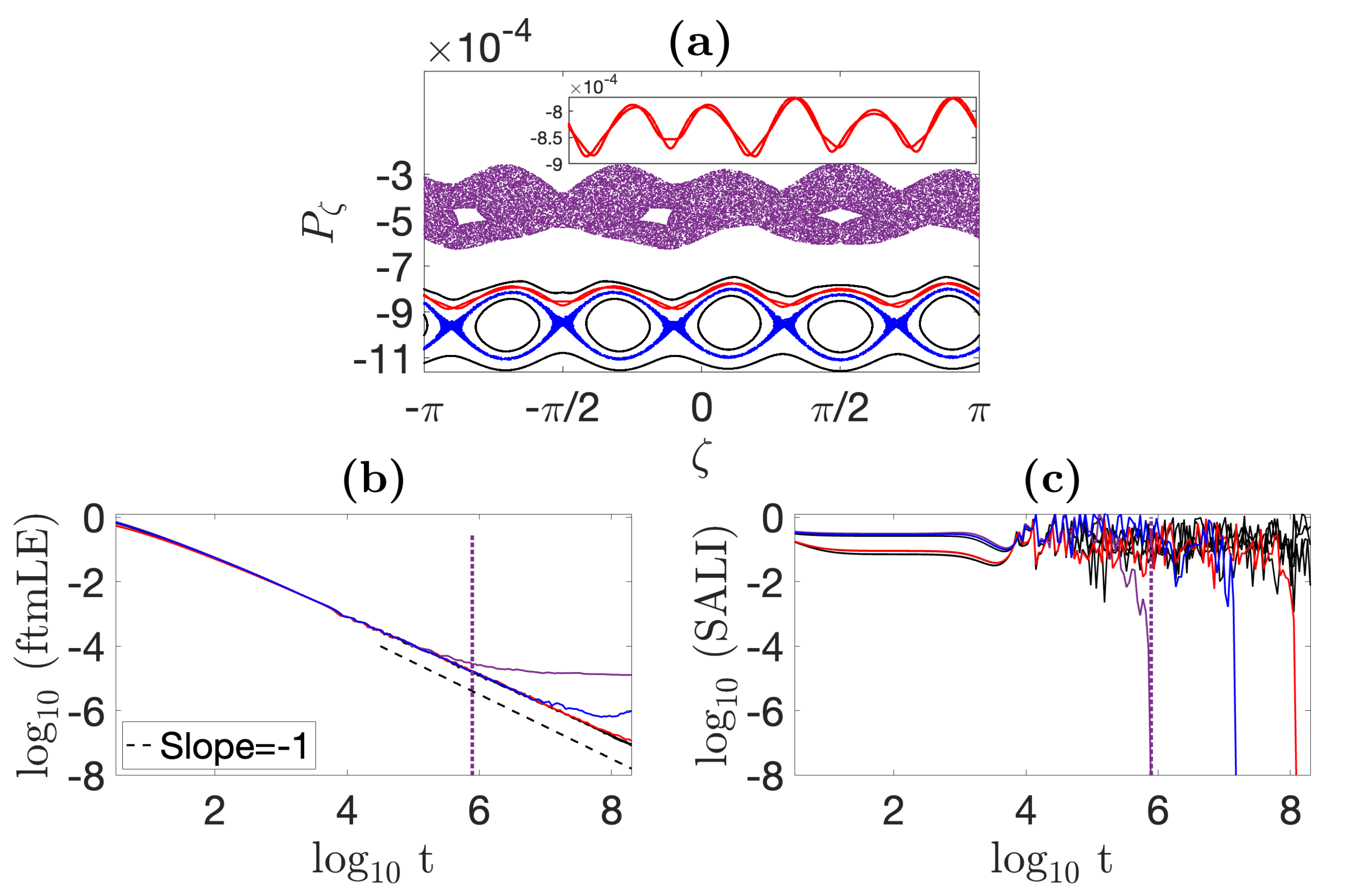}
	\caption{(a) Poincar{\'e} surface of section $(\theta=0, \mod 2\pi)$ of guiding center orbits [Eq.~\eqref{eq:GCM-Ham-q=1}] with normalized energy $E = 8.131\times10^{-6}$ and magnetic moment $\mu = 8.1423\times 10^{-6}$ for $P_{\zeta0} = -0.85 \times 10^{-3}$ (blue), $P_{\zeta0} = -0.81 \times 10^{-3} $ (red), $P_{\zeta0} = -0.36 \times 10^{-3} $ (purple). The perturbing mode numbers are  $n_{1,2} =5, 3$, $m_{1,2} =1,1$. Regular orbits are depicted in black, whereas chaotic orbits are depicted in purple (scattered points) and blue/red (at the borders of island chains). [Inset: Zoomed in weakly chaotic red-colored orbit. This orbit has to be integrated until the final time of $t=10^9$ to show its chaotic nature]. Time evolution of the ftmLEs (b) and the SALI (c) for the orbits in (a). The dashed line in (b) corresponds to the slope of $-1$, while the purple dotted vertical line in (b) and (c) indicates $\log_{10}(t)=5.894$ (see text for more details).}
	\label{fig:Fig1}
\end{figure}

Characteristic cases of regular and chaotic GC orbits are shown in a Poincar{\'e} surface of section in Fig.~\ref{fig:Fig1}(a), where resonant island chains surrounded by chaotic regions are shown. It is worth emphasizing that under the consideration of a constant safety factor $q=1$, the locations of the kinetic resonances are determined solely by the particle orbital frequencies (bounce/transit poloidal and averaged toroidal precession) whose ratio defines the orbit helicity profile as a function of $P_\zeta$, as will also be discussed in Section~\ref{sec:Magnetic versus Kinetic Chaos in Toroidal Plasmas}.

Their respective ftmLE and SALI are respectively depicted in Fig.~\ref{fig:Fig1} (b) and (c). For the considered regular orbits (black curves/points) the ftmLE tends to zero following a power law proportional to $t^{-1}$, while it eventually saturates to a positive number for the chaotic orbits. The ftmLE saturates to a constant value at different times depending on the strength of chaos: the ftmLE of the orbit colored in purple saturates to its limiting value, $\sigma$ $\approx 1.277 \times 10^{-5}$, at time $t\approx10^6$. This orbit is more chaotic than the orbit colored in blue whose ftMLE saturates to $\sigma$ $\approx 9.686 \times 10^{-7}$ at $t\approx4\times10^7$, whereas the curve corresponding to the red-colored orbit has not saturated yet until the considered final time $2\times10^8$ and is still following the $t^{-1}$ power law (slope denoted by dashed line in Fig.~\ref{fig:Fig1}(b)). On the other hand, the SALI remains practically constant in time for regular orbits [black curve in Fig.~\ref{fig:Fig1}(c)], while it eventually tends to zero exponentially fast for the studied chaotic orbits. The SALI manages to clearly identify the chaotic nature of the three studied chaotic orbits faster and more accurately than the ftmLE because the introduction of chaos immediately results in SALI values which differ by many orders of magnitude from those observed for regular orbits. On the other hand, although the ftmLE of the chaotic orbits will gradually start showing deviations from the $t^{-1}$ decay observed for regular orbits, its values will not be very different between the two cases to allow the clear discrimination between them by only registering the index values, without having to check by eye its evolution. Relying only on the numerical value of a chaos indicator is of utmost practical importance when studying the global behavior of a system by investigating ensembles of many orbits. To illustrate these value differences we consider the purple-colored orbit of Fig.~\ref{fig:Fig1} and compare its SALI and ftmLE values to the ones of the regular, black colored orbit in the same figure. For $\log_{10} t \approx 5.894$ [this time is indicated by a purple dotted vertical line in Fig.~\ref{fig:Fig1}(b) and (c)] we have $\log_{10}(\mbox{SALI}) \approx -8.696$ [$\log_{10}(\mbox{SALI}) \approx -1.520$] and $\log_{10}(\mbox{ftmLE})  \approx -4.560$ [$\log_{10}(\mbox{ftmLE})  \approx -4.762$] for the chaotic [regular] orbit. Furthermore, the strength of chaos is reflected on the time required for SALI to become zero, i.e. less time corresponds to larger mLE values. It is worth emphasizing that the orbit corresponding to the red curve could be mistakenly considered as a regular one, by just inspecting the time evolution of its ftmLE, which practically remains proportional to $t^{-1}$ within the considered computation time [red curve in Fig.\ref{fig:Fig1}(b)], whereas SALI clearly reveals the true nature of this weakly chaotic orbit, in the same time frame [red curve in Fig.\ref{fig:Fig1}(c)]. It should be mentioned that the practical implications of weak chaos can be assessed in comparison to other relevant time scales such as the time scales associated with particle scattering due to Coulomb collisions or microturbulence.

\begin{figure}
\centering
\includegraphics[width=1\textwidth,keepaspectratio]{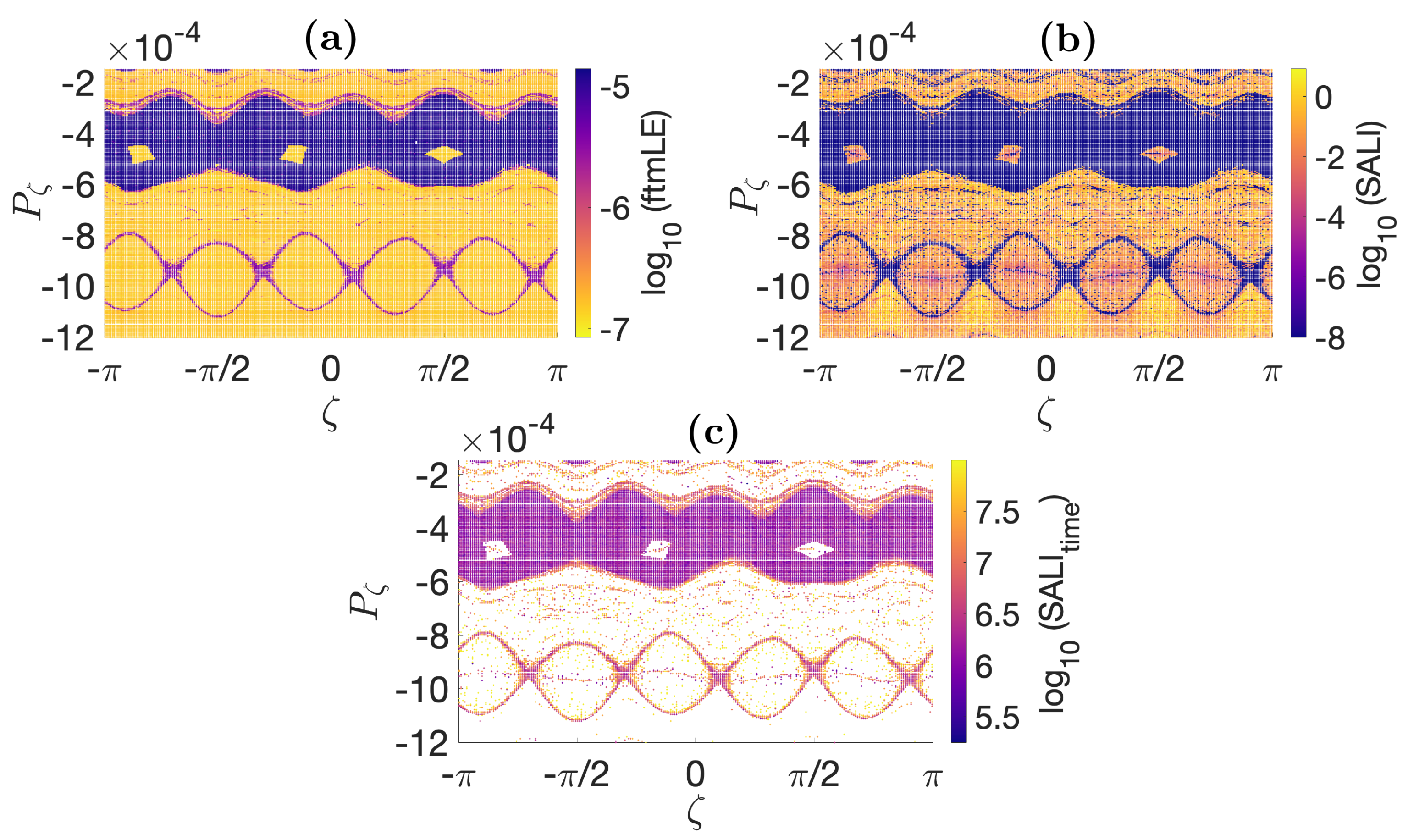}
\caption{Poincar{\'e} surface of section $(\theta=0, \mod 2\pi)$ of guiding center orbits coloured according to the ftmLE (a) and SALI (b) values after $10^8$ integration time units. The time required for SALI to detect chaos is depicted in (c). Initial conditions are uniformly distributed in  $(\zeta,P_\zeta) \in [-\pi, \pi]\times[-1.2 \times 10^{-3}, -1.5\times 10^{-4}]$. The values of normalized energy $E$ and magnetic moment $\mu$ are the same with Fig.~(\ref{fig:Fig1}).}
\label{fig:Fig2}
\end{figure}

The Poincar{\'e} surface of sections of GC orbits coloured according to the ftmLE and SALI values after $10^8$ integration time units, for a dense set of uniformly distributed initial conditions, are respectivelly shown in Figs.~\ref{fig:Fig2}(a) and (b). Generally, in both panels, yellow regions represent regular motion, dark blue areas represent chaotic motion, and in-between colors represent weakly chaotic orbits requiring more integration time to reveal their chaotic nature. It is worth noting that the characterization of orbits as regular (yellow-colored points) in both Figs.~\ref{fig:Fig2}(a) and (b), is based on results until the final integration time considered ($10^8$ time units). Some of these orbits might actually be weakly chaotic but their true nature could be revealed only at longer times, when their ftmLE will saturate to a small positive value and their SALI will abruptly drop to zero.

From the comparison of Figs.~\ref{fig:Fig2}(a) and (b), it is clearly seen that SALI provides significantly higher resolution analysis of the phase space chaoticity by identifying narrow chaotic regions related to separatrices of higher-order resonant island chains, which are not clearly seen in Fig.~\ref{fig:Fig2}(a). This is an important advantage of the SALI method. The time required for SALI to detect chaos, i.e. the time at which SALI becomes smaller than $10^{-8}$, is depicted in Fig.~\ref{fig:Fig2}(c), where the white areas correspond to regions for which SALI$>10^{-8}$ until the end of the considered integration time. It is worth noting that the color plots produced by the SALI method are not only more accurate than the ones created
through the evaluation of the ftmLE in correctly capturing the (even weakly) chaotic nature of orbits, but their computation is also quite efficient. In particular, the SALI plot of Fig.~\ref{fig:Fig2}(b)
requires only about $5\%$ more CPU time than Fig.~\ref{fig:Fig2}(a), despite the fact that the SALI computation requires the time evolution of two deviation vectors instead of just one needed for the estimation of the mLE. This is due to the fact that the evolution of orbits is stopped when their SALI becomes smaller than $10^{-8}$.

\begin{figure}
        \centering
	\includegraphics[width=1\textwidth,keepaspectratio]{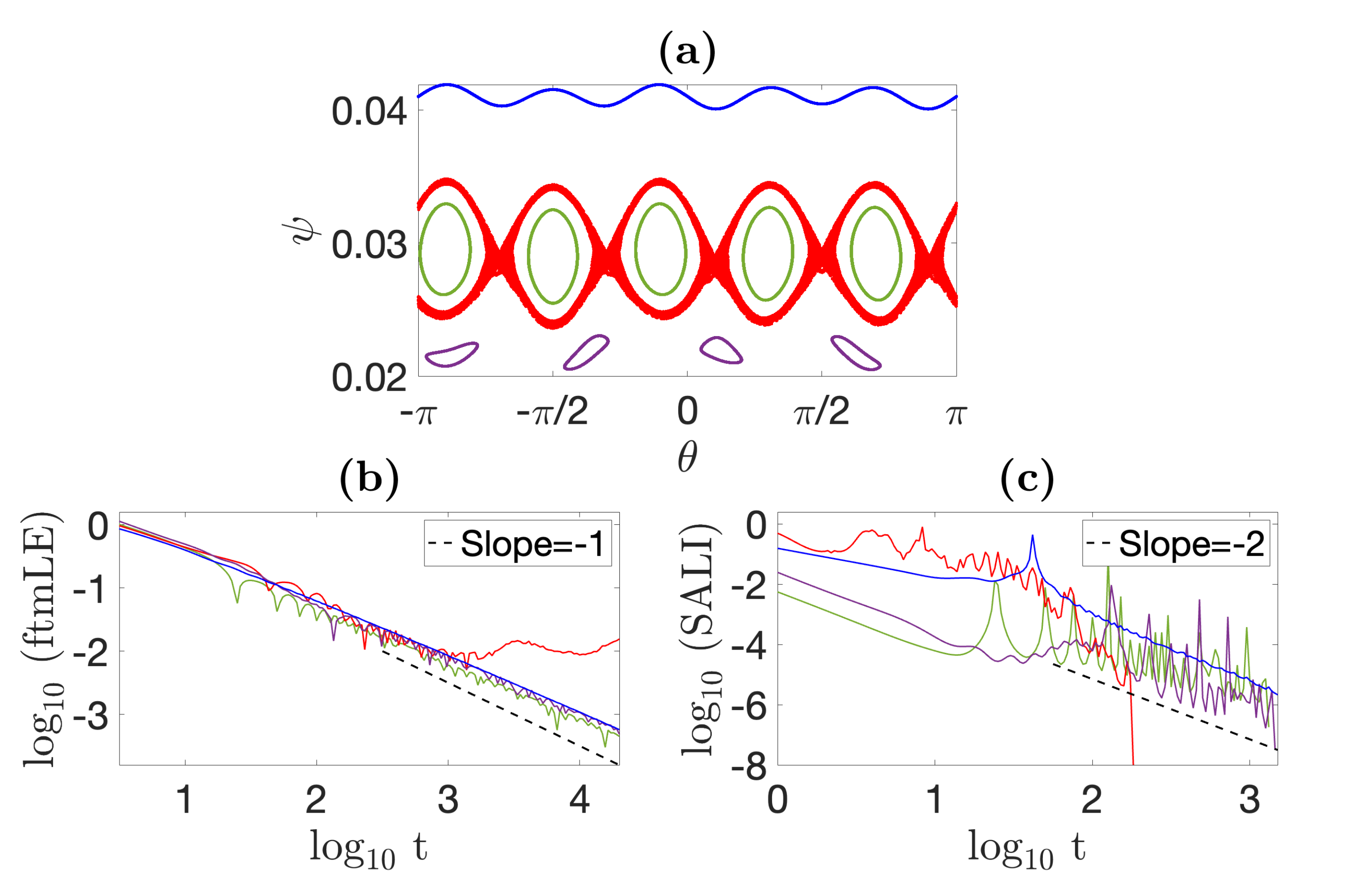}
	\caption{(a) Poincar{\'e} surface of section $(\zeta=0, \mod 2\pi)$ of magnetic field lines [Eq.~\eqref{eq:HAM-MFLInteg}] for three different regular orbits with initial conditions  $(0.12, 0.0212)$ (purple),  $(0.03, 0.03)$ (green) and $(0., 0.04)$ (blue) and a chaotic orbit with initial $(0.032, 0.033)$ (red). The mode numbers are the same as those in Fig. \ref{fig:Fig1}. Time evolution of the ftmLEs (b) and the SALI (c) for all of the orbits in (a). The dashed lines in (b) and (c) correspond to the functions proportional to $t^{-1}$ and $t^{-2}$, respectively.}
\label{fig:Fig3}
\end{figure}
 
A similar analysis is performed for the case of magnetic field lines described by the Hamiltonian equations given in Eq.~\eqref{per Ham MFL} and a safety factor given by Eq.~\eqref{gen q factor}. Characteristic cases of regular and chaotic field lines are depicted in a Poincar{\'e} surface of section in Fig.~\ref{fig:Fig3}(a), along with the time evolution of the ftmLE and the SALI in panels (b) and (c), respectively. For the regular orbits the ftmLE tends to zero following a power law $t^{-1}$ whereas it becomes constant for the chaotic orbit. On the other hand, the SALI tends to zero following a power law $t^{-2}$ for regular orbits, in agreement to the theoretical prediction of Eq.~\eqref{Prop:SALI-SM}, as in this case the system's phase space is 2D, while it tends exponentially fast to zero for the considered chaotic orbit (red points/curves in Fig.~\ref{fig:Fig3}). From the results of Figs.~\ref{fig:Fig3}(b) and (c), it can clearly be seen that SALI distinguishes the chaotic orbit very fast at time $t\approx180$.

\section{\label{sec:Magnetic versus Kinetic Chaos in Toroidal Plasmas}Magnetic versus Kinetic Chaos in Toroidal Plasmas}

In this section, we explore and systematically compare magnetic versus kinetic chaos for particle orbits corresponding to the entire range from thermal to energetic particles. Magnetic flux surfaces break into resonant island chains at regions $\psi_{res}$ where the resonance condition
\begin{equation}
    q(\psi_{res})=m/n,
\end{equation}
is met, with $m,n$ the corresponding perturbation mode numbers.
The presence of non-axisymmetric perturbations also modify the topology of the GC phase space, with resonant islands formed at positions where the kinetic-$q$ factor ($q_{kin}$), defined as the ratio of the bounce/transit-averaged toroidal precession frequency to the bounce/transit frequency \cite{biewer2003electron, Gobbin2008,  White2015, zestanakis2016orbital, shinohara2018estimation, Bierwage2023, Anastassiou2023} fulfills the condition $q_{kin}=m/n$. The $q_{kin}$ depends on the kinetic characteristics of the particle through the three COM. As intuitively expected, low-energy particles closely follow the magnetic field lines and have a $q_{kin}$ close to $q$, so their chaoticity, related to resonant island separatrices and/or resonant overlap, is strongly related to that of the magnetic field lines. However, for more energetic particles, the locations of resonant islands and conditions for extended chaos in the GC phase space, are radically different from those of the magnetic field.

In the following analysis, we consider a typical $q$ factor profile given by Eq.~\eqref{gen q factor} with parameters $\nu=2$, $q_{ma}=1.1$, $q_{w} = 4.0$ and $\psi_{w} = 0.05$. We investigate the case of two perturbation modes with mode numbers $(m,n)=(3,2)$ and $(5,2)$.
For simplicity we neglect the radial profile of the perturbations so that the coefficients $\alpha_{m,n}$ are considered independent of $\psi$ and of equal values $\alpha_{3,2}=\alpha_{5,2}=\epsilon=7.5\times 10^{-5}$. Moreover, in the following cases, we consider hydrogen particles in a magnetic field of $B_0=1$ T. 

\begin{figure}
	\centering
	\includegraphics[width=1\textwidth,keepaspectratio]{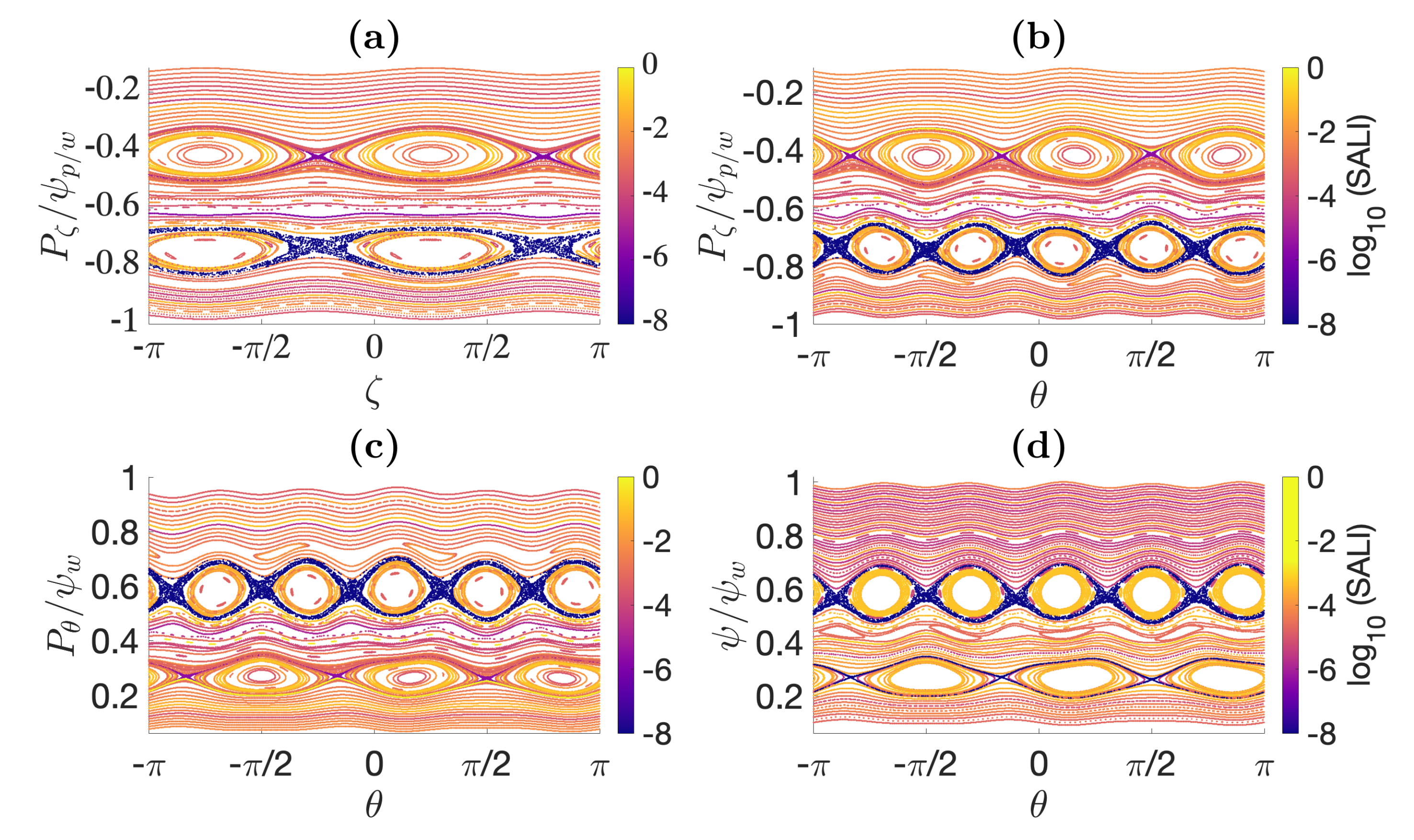}
	\caption{(a), (b) and (c) Kinetic Poincar{\'e} surfaces of section for low-energy particles with $E=3.4$eV and $\mu B_{0}=2.6$eV. (d) Magnetic Poincar{\'e} surface of section. All orbits are coloured according to their SALI value calculated for $10^{8}$ time units. Two perturbation modes with mode numbers $(m,n)=(3,2), (5,2)$ and equal amplitudes are considered $\alpha_{3,2}=\alpha_{5,2}=\epsilon=7.5\times 10^{-5}$. }
\label{fig:Fig4}
\end{figure}

The case of low-energy particles with $E=3.4$eV and $\mu B_{0}=2.6$eV is depicted in Fig.~\ref{fig:Fig4}. Kinetic Poincar{\'e} surfaces of section $(P_\zeta, \zeta)$ and $(P_\zeta, \theta)$ are shown in Figs. \ref{fig:Fig4}(a) and (b), with the orbits being coloured according to their final SALI value. The number of islands in each island chain is determined by the respective mode numbers. The kinetic Poincar{\'e} surface of section $(P_\theta, \theta)$ is shown in  Fig. \ref{fig:Fig4}(c). Since $P_\theta=\psi$ (LAR approximation) this kinetic Poincar{\'e} surface of section can be directly compared to the magnetic one shown in Fig. \ref{fig:Fig4}(d). The comparison clearly shows that kinetic and magnetic chaos appear close to the separatrices of the island chains and confirms that chaos is located in the same radial ($\psi$) position, with the degree of chaoticity, as quantified by SALI, being similar. 


\begin{figure}
	\centering
	\includegraphics[width=1\textwidth,keepaspectratio]{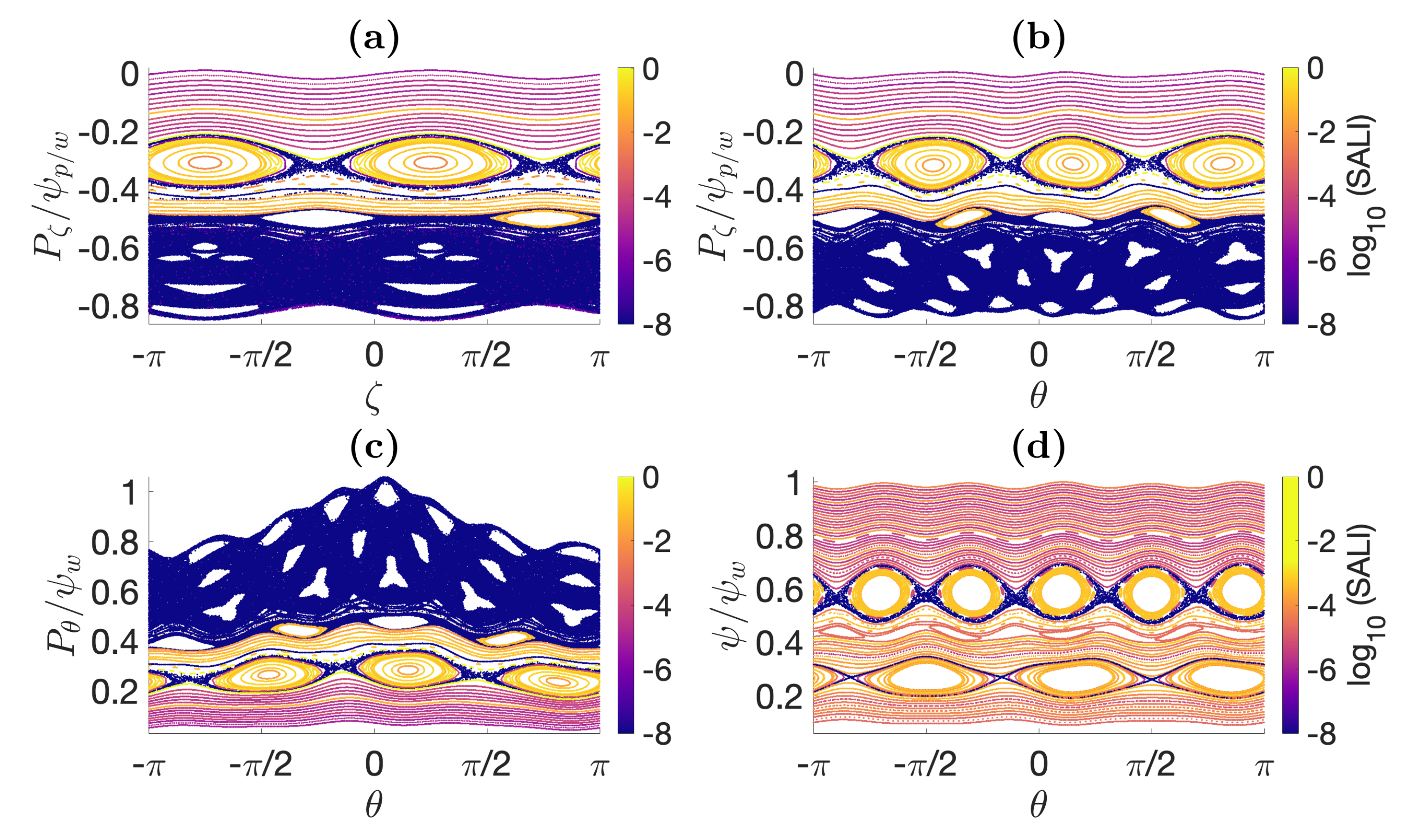}
	\caption{(a), (b) and (c) Kinetic Poincar{\'e} surfaces of section for thermal particles with $E=2.9$keV and $\mu B_{0}=2.0$keV. (d) Magnetic Poincar{\'e} surface of section. All orbits are coloured according to their SALI value calculated for $10^{8}$ time units. Two perturbation modes with mode numbers $(m,n)=(3,2), (5,2)$ and equal amplitudes are considered $\alpha_{3,2}=\alpha_{5,2}=\epsilon=7.5\times 10^{-5}$. }
 \label{fig:Fig5} 
\end{figure}

The case of thermal particles with $E=2.9$keV and $\mu B_{0}=2.0$keV is depicted in Fig.~\ref{fig:Fig5}. As shown in the kinetic Poincar{\'e} surfaces of section in Figs. \ref{fig:Fig5}(a) and (b), the resonant islands appear in different values of $P_\zeta$ and the lower chaotic region is significantly more extended in comparison to Figs. \ref{fig:Fig4}(a) and (b). Moreover, as shown in Figs. \ref{fig:Fig5}(c) and (d), the kinetic chaos related to the upper primary resonant island chain extends to a significantly larger range of radial positions in comparison to magnetic chaos. In fact, this chaotic region connects to the wall ($P_\theta / \psi_w =1$), suggesting perturbation-induced stochastic particle loss. Also, secondary island chains appear within the chaotic regions of the kinetic Poincar{\'e} surface of section.        

\begin{figure}
    \centering
    \includegraphics[width=1\textwidth,keepaspectratio]{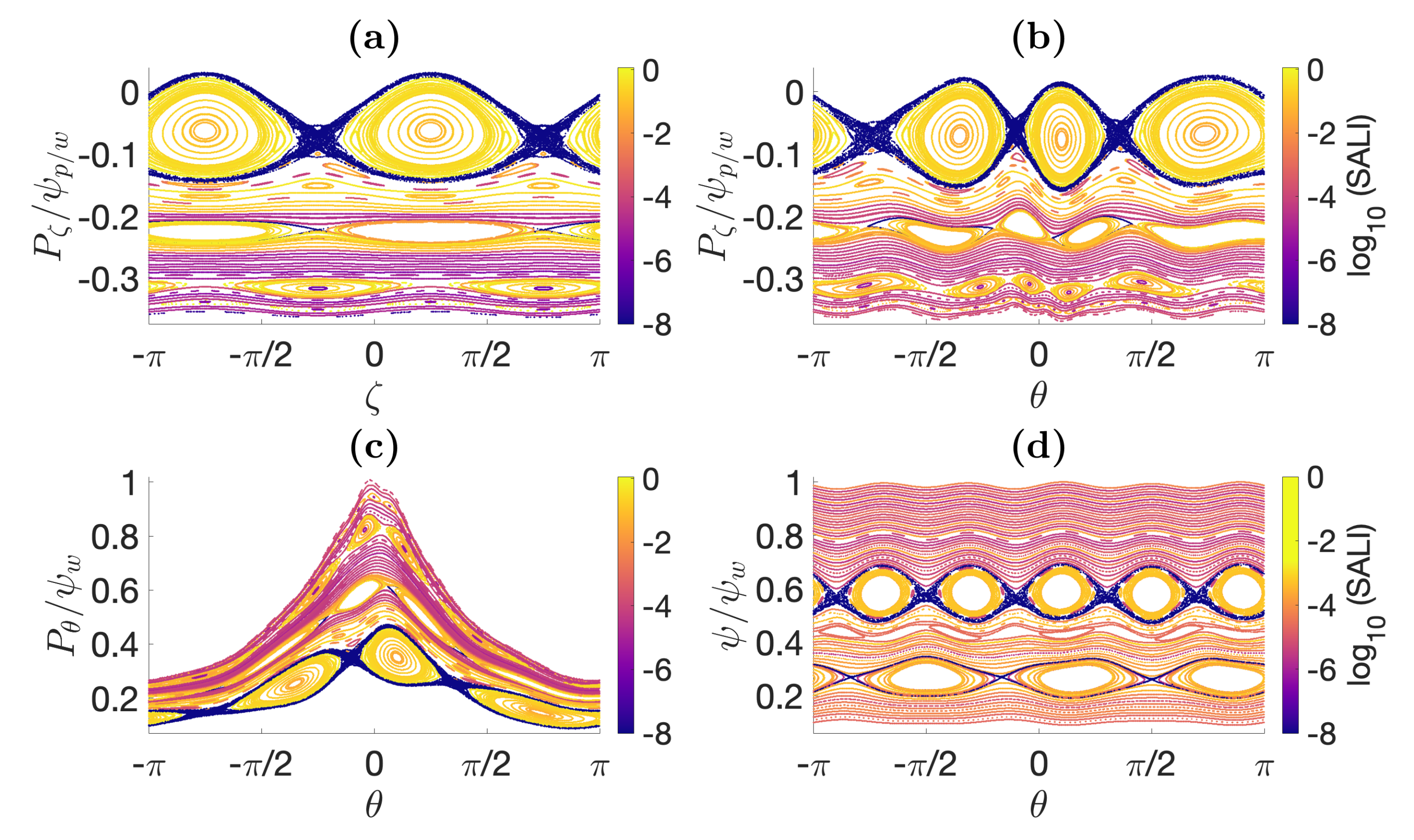}
    \caption{(a), (b) and (c) Kinetic Poincar{\'e} surfaces of section for energetic particles with $E=39.1$keV and $\mu B_{0}=33.1$keV. (d) Magnetic Poincar{\'e} surface of section. All orbits are coloured according to their SALI value calculated for $10^{8}$ time units. Two perturbation modes with mode numbers $(m,n)=(3,2), (5,2)$ and equal amplitudes are considered $\alpha_{3,2}=\alpha_{5,2}=\epsilon=7.5\times 10^{-5}$. }
    \label{fig:Fig6}
\end{figure}

The case of higher-energy particles with $E=39.1$keV and $\mu B_{0}=33.1$keV is depicted in Fig.~\ref{fig:Fig6}, where the kinetic Poincar{\'e} surfaces of section clearly show that the resonance chains are located at quite different values of $P_\zeta$ in comparison to the previous lower-energy particles \cite{zestanakis2016orbital, antonenas2021analytical}. In this case, the extent of kinetic chaos is markedly reduced in comparison to the case of thermal particles with $E=2.9$keV and $\mu B_{0}=2.0$keV (Fig.~\ref{fig:Fig5}) and it is located close to the separatrices of the primary island chains. 

The kinetic and magnetic Poincar{\'e} surfaces of section in Figs.~\ref{fig:Fig4}-\ref{fig:Fig6} clearly show that the same magnetic perturbations, and therefore the same magnetic field chaos, has completely different effects on particles with different kinetic characteristics. Although, Poincar{\'e} surfaces of section can serve for illustrating characteristic cases, a systematic global overview of the effect of the magnetic field chaos, due to a specific set of modes, on the entire range of particles can be provided in the space of the constants of the motion uniquely labeling each GC orbit and characterizing it in terms of being trapped or passing, and confined or lost. Moreover, such a description necessitates chaos quantification enabling the assignment of a chaos measure at each point (orbit) in the space of the constants of the motion. 

\begin{figure}
	\centering
 	\includegraphics[width=0.475\textwidth,keepaspectratio]{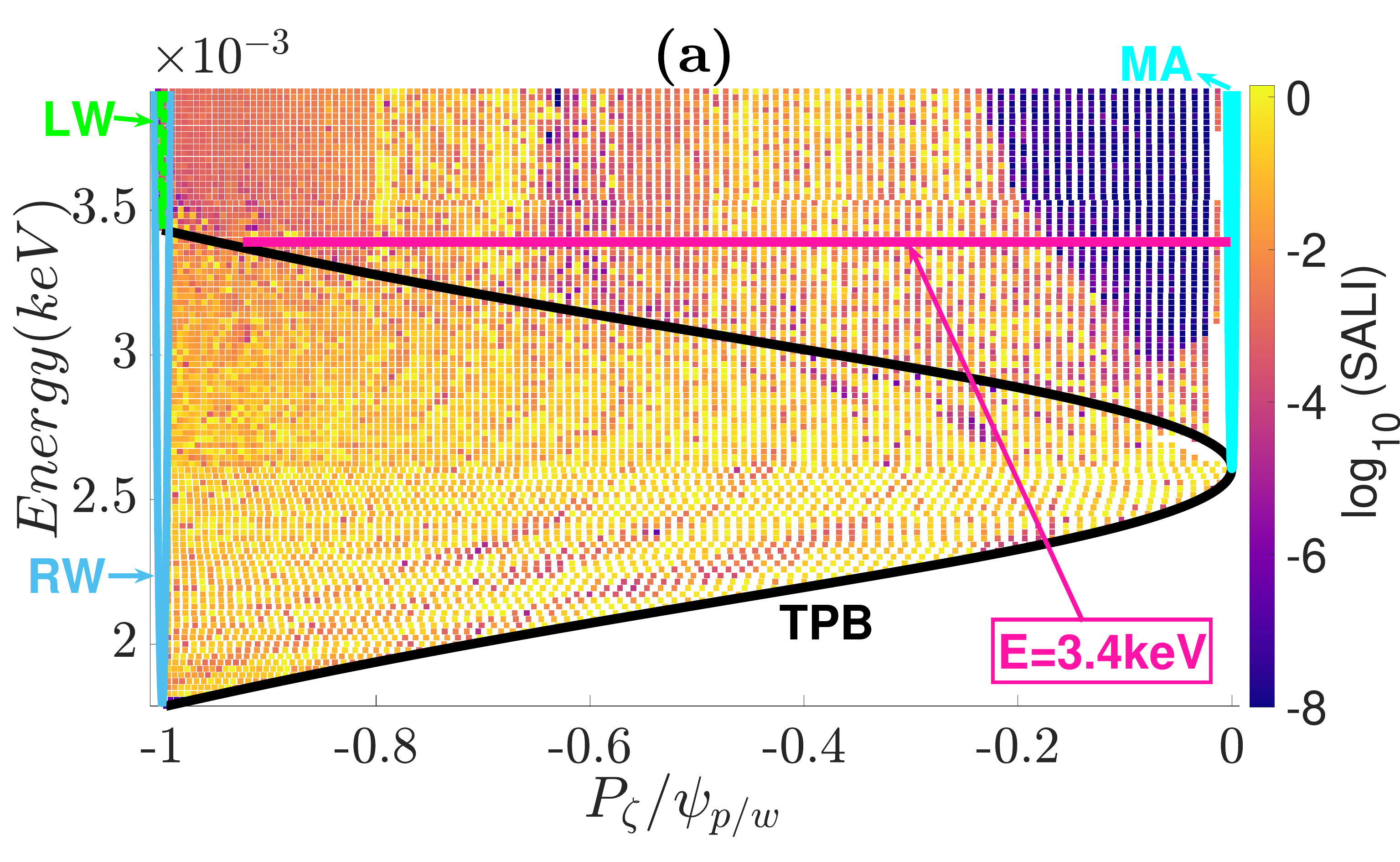}
      \includegraphics[width=0.475\textwidth,keepaspectratio]{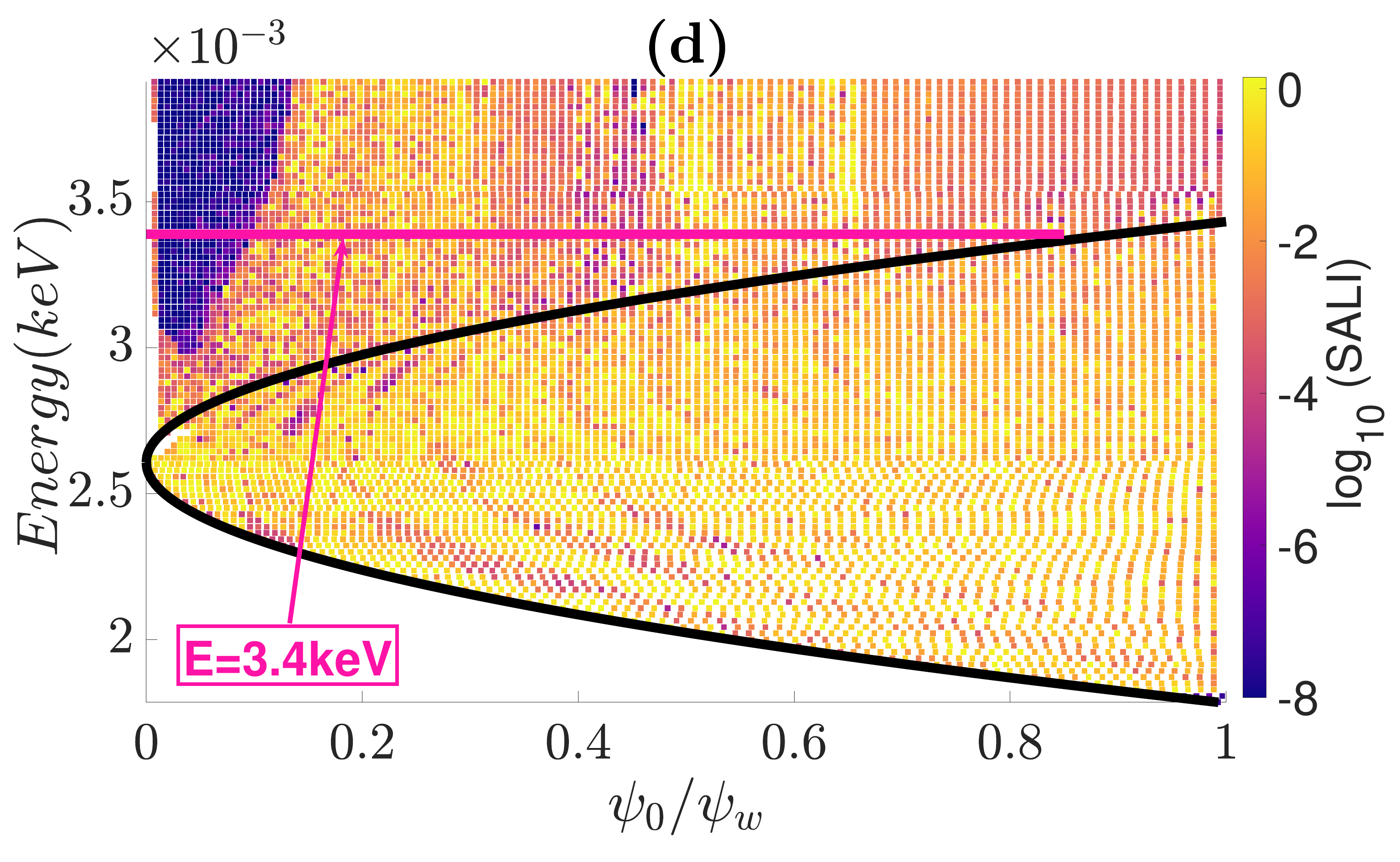}
   \includegraphics[width=0.475\textwidth,keepaspectratio]{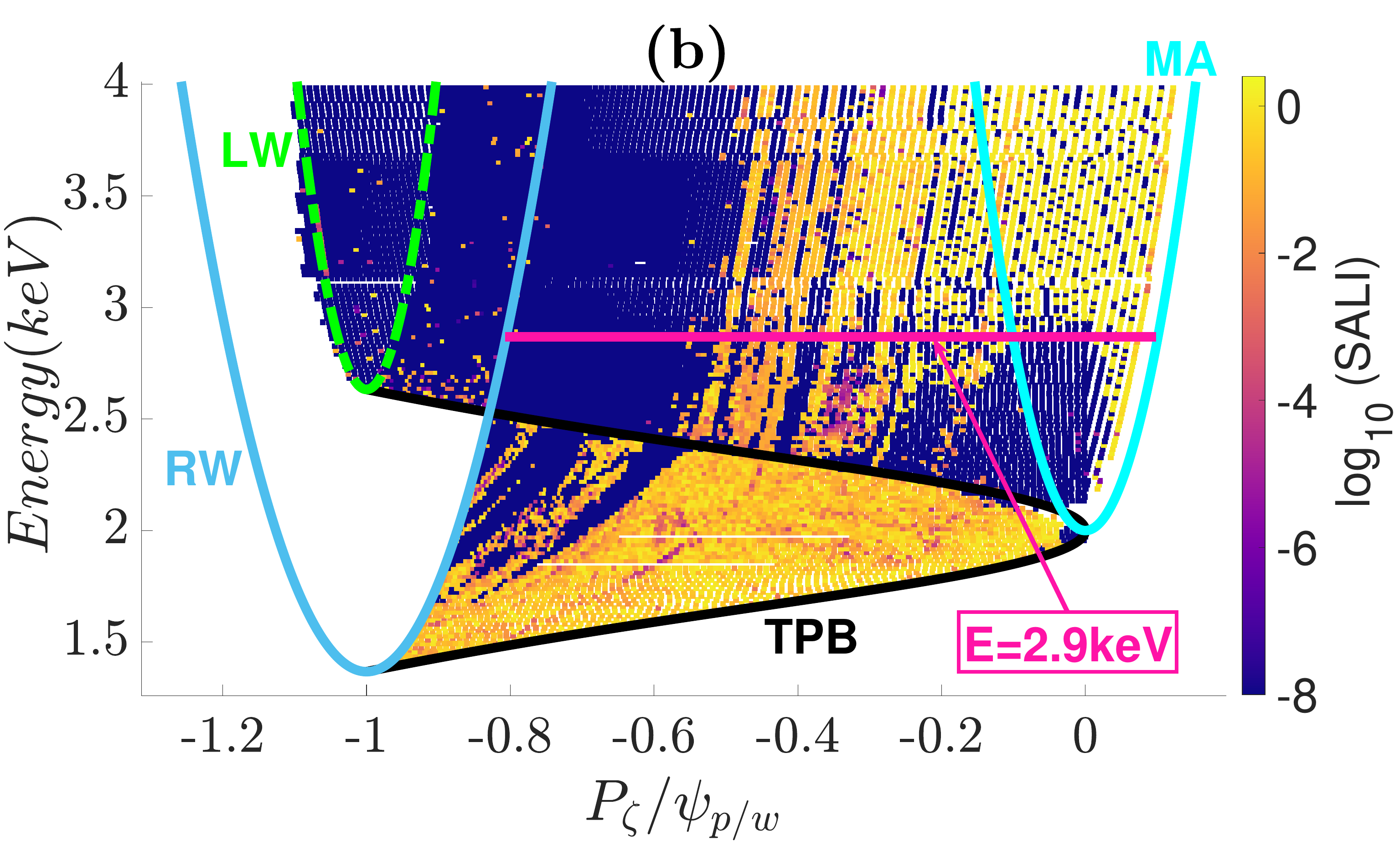}   
       \includegraphics[width=0.475\textwidth,keepaspectratio]{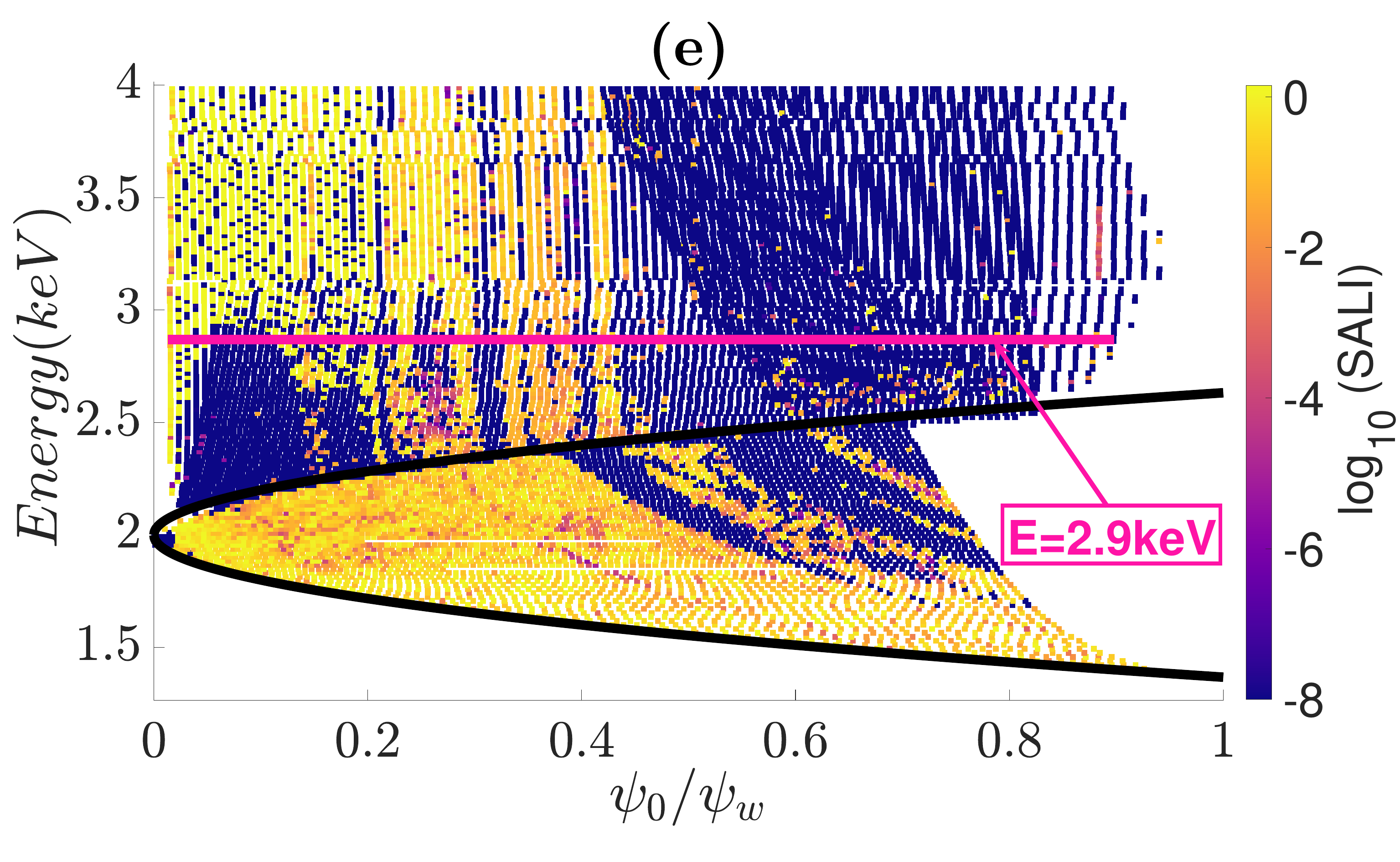}
 	  \includegraphics[width=0.475\textwidth,keepaspectratio]{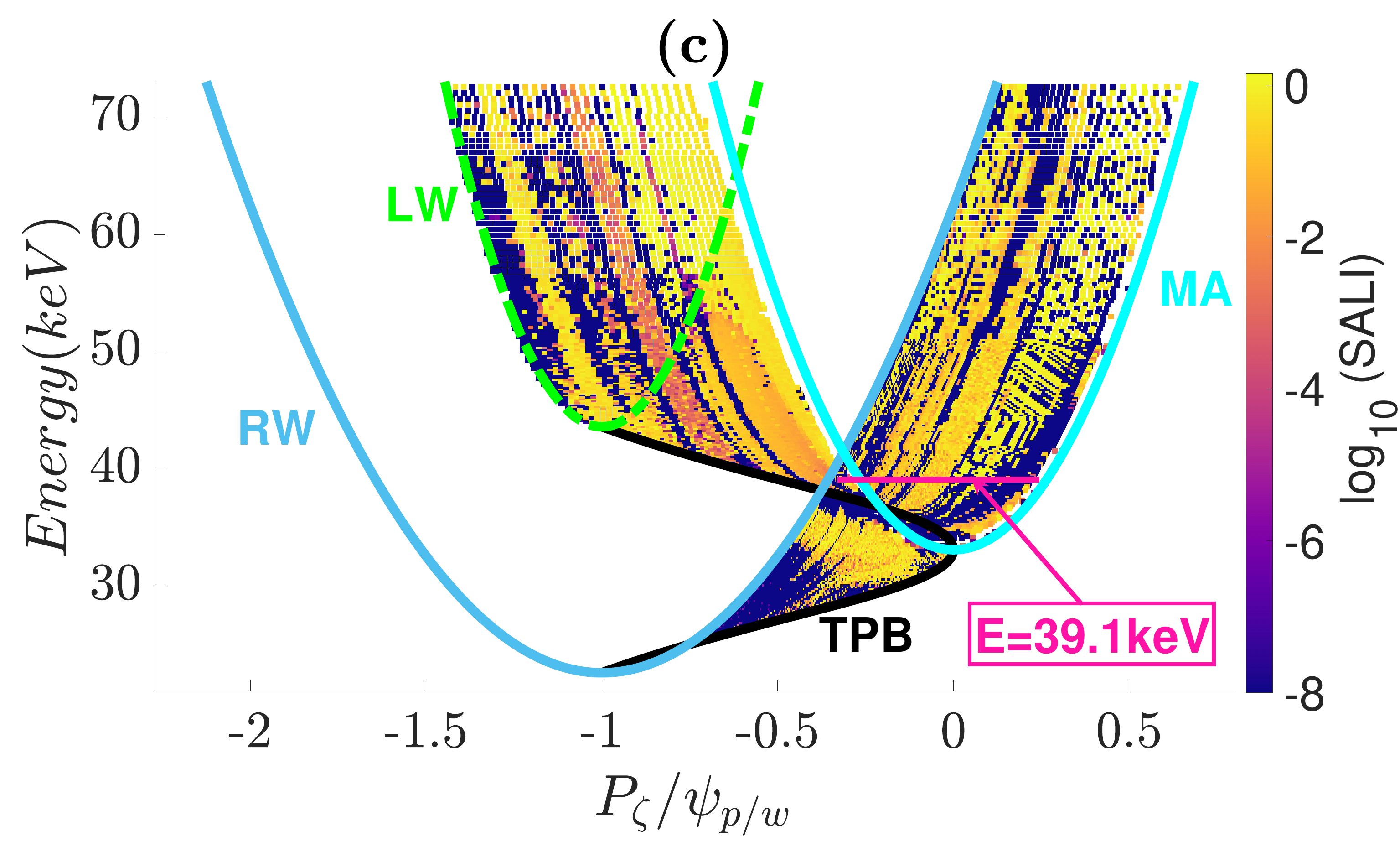}
    \includegraphics[width=0.475\textwidth,keepaspectratio]{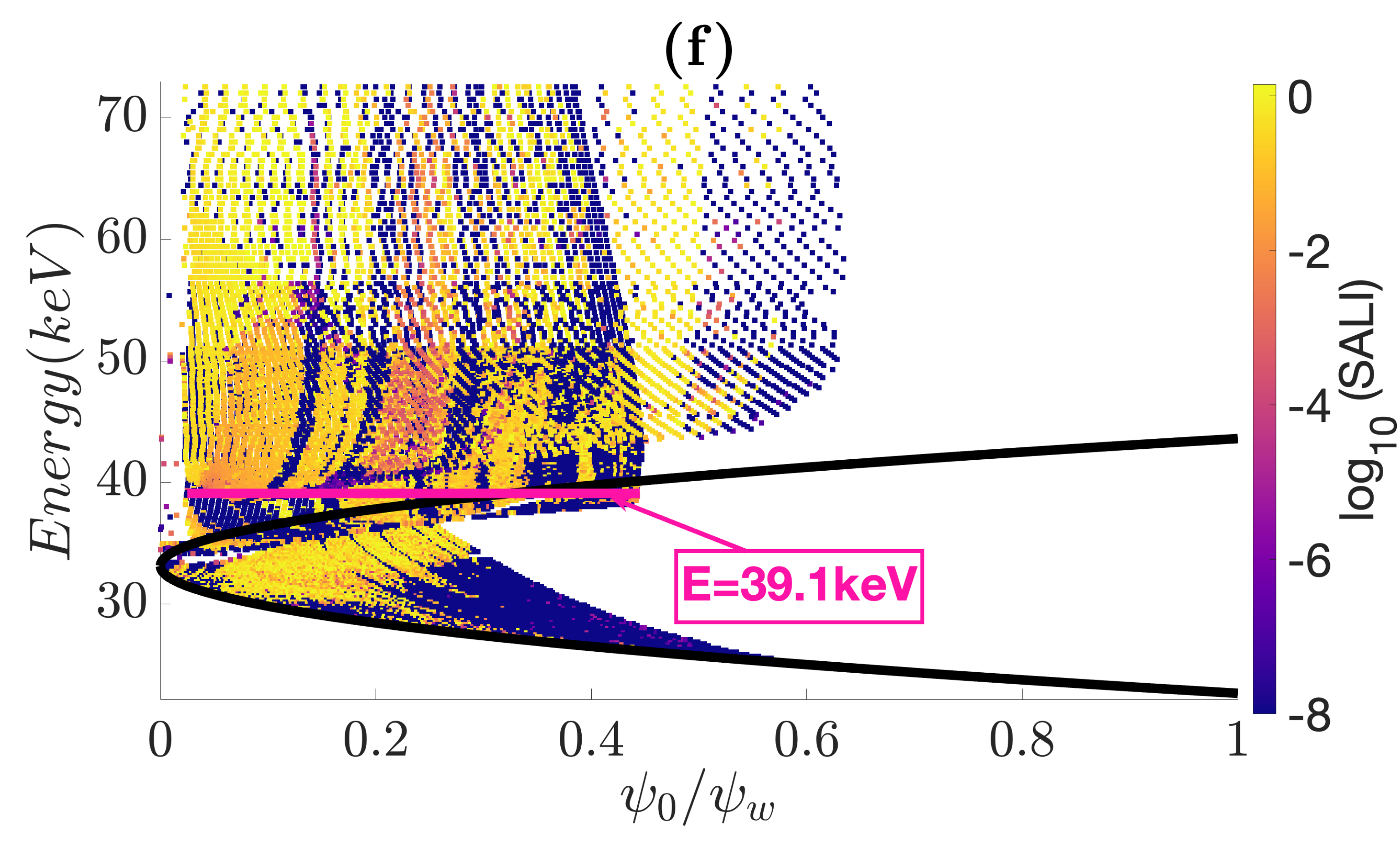}
    
	\caption{Constant $\mu$ plane cuts of the 3D constants-of-the-motion space  $(E,P_\zeta,\mu)$ (a-c) or $(E,\psi_0,\mu)$ (d-f) with $\mu B_0=2.6$eV (a, d), $\mu B_0=2.0$keV (b, e), and $\mu B_0=33.1$keV (c, f) corresponding to perturbation modes as in Figs. \ref{fig:Fig4}-\ref{fig:Fig6}. Each point defines an orbit and it is colored according to the calculated SALI value, with dark blue and yellow areas representing chaotic and regular motion, respectively. The specific energy level considered in the Poincar{\'e} surfaces of section in Figs. \ref{fig:Fig4}-\ref{fig:Fig6} is denoted by a horizontal pink line. The right wall (RW), left wall (LW), magnetic axis (MA), and trapped-passing boundary (TPB) [Eqs. \eqref{walls}, \eqref{magnetic axis},  \eqref{trapped passing boundary}] are depicted by solid light blue, dashed light green and solid cyan parabolas respectively, and provide an overview of the specific kinetic characteristics of particle orbits with different chaoticity. The radial position where kinetic chaos occurs can be obtained in (d-f) where the reference flux surface $\psi_0$ of its orbit is used instead of $P_\zeta$ according to Eq.~(\ref{psi0}). 
 }
	\label{fig:Fig7}
\end{figure}

The three-dimensional COM space $(E, P_\zeta, \mu)$ is systematically dissected in Fig.~\ref{fig:Fig7} (a-c), where the SALI value, calculated at each point, quantifies the chaoticity of the respective orbit. Moreover, these diagrams reveal whether the chaotic orbits correspond to trapped or passing particles as well as their positions with respect to the magnetic axis and the wall, as defined by Eqs. \eqref{walls}, \eqref{magnetic axis},  \eqref{trapped passing boundary}. It is evident that the same magnetic chaoticity (same perturbative modes) corresponds to completely different kinetic chaoticity depending strongly on the kinetic characteristics of the particles. The diagrams provide a clear and detailed overview of the effect of a specific set of perturbative modes on the chaotic particle, energy and momentum transport and confinement in toroidal fusion devices. Narrow chaotic strips correspond to localized chaos at separatrices of isolated resonant island chains, as illustrated in Figs. \ref{fig:Fig4}-\ref{fig:Fig6}, whereas wider chaotic regions correspond to extended chaos due to resonance overlap. The relative position of the chaotic areas with respect to the position of the wall and the trapped/passing boundary, provides information on  particle losses as well as complex particle trapping-detrapping dynamics due to the formation of a chaotic sea connecting both sides of the separatrix (trapped/passing boundary) of the unperturbed particle motion. It is worth noting the remarkable high degree of aggregation of the information presented here, in comparison to Poincar{\'e} surfaces of section [Figs. \ref{fig:Fig4}-\ref{fig:Fig6}], each one corresponding to a constant energy line, as well as the importance of efficiently quantifying chaos in terms of SALI that enables the assignment of a chaos measure to each orbit corresponding to a point of the dense grid used to depict the fine details of kinetic chaos. 

The radial position and extent of kinetic chaos cannot be rigorously defined due to the fact that even the unperturbed particle orbits, especially the more energetic ones, are not restricted on a single flux surface due to GC drifts. However, a flux surface of reference $\psi_0$, around which the particle drifts, can be defined for each unperturbed orbit in order to have an estimation for the approximate radial distance where kinetic chaos actually takes place. This flux surface $\psi_0$ is defined according to  
\begin{equation}
    \psi_p(\psi_0)=\left< \psi_p (\psi) \right >=\left<\frac{\pm g(\psi)}{B(\psi,\theta)}\sqrt{2\left(E-\mu B(\psi,\theta)\right)} \right> - P_\zeta,
\end{equation}
where the brackets denote orbit averaging. For trapped particles, it suffices to consider $\psi_0$ as the value of $\psi$ at the points of velocity reversal, where the trapped particles spend the most of their time. To lowest order in a LAR configuration, $\psi_0$ can be defined for all types of orbits, in terms of the constant variables $(E, \mu, P_\zeta)$, as follows 
\begin{equation}
    \psi_{p}(\psi_{0}) = \left\{
    \begin{array}{ll}
     -P_\zeta    & \text{, trapped}\\
     \pm \sqrt{2(E-\mu)}-P_\zeta   & \text{, co/counter-passing}
    \end{array}
    \right.
    \label{psi0}
\end{equation}
 and can be used as an alternative to $P_\zeta$ constant of the motion uniquely labeling each orbit along with $E$ and $\mu$ and characterizing the radial position and extent of kinetic resonances, islands and chaos \cite{pinches1998hagis, antonenas2021analytical} (for other alternative sets of COM see Refs. \cite{Bierwage2022b, Benjamin2023}). The respective information of the kinetic chaoticity in the $(E,\psi_0,\mu)$ space is depicted in Figs.~\ref{fig:Fig7}(d-f).

\clearpage

\section{\label{sec:Summary and Conclusion}Summary and Conclusion}
The presence of non-axisymmetric perturbations results in non-integrability of the Hamiltonian systems describing magnetic field lines and charged particle motion in toroidal plasmas and therefore plays a crucial role in particle, energy and momentum transport and confinement properties of fusion devices. Low-energy particles closely follow magnetic field lines and have approximately the same chaoticity with them. The effect of the non-axisymmetric perturbation on the topology of the magnetic field lines is localized to regions where magnetic resonant islands or chaotic regions, under conditions for resonance overlap, are formed. Higher-energy particles undergo large drifts across the magnetic field lines, and the effect of the non-axisymmetric perturbations on their orbits is radically different from the effect of the same perturbations on the magnetic field lines. In this general case, the locations of kinetic and magnetic resonant island chains, the conditions for kinetic and magnetic chaos, as well as the degrees of kinetic and magnetic chaoticities are drastically different. Chaotic transport of high-energy particles determines the confinement limitations and the performance of the fusion devices and necessitates the understanding of the relation between magnetic and kinetic resonances and the respective magnetic and kinetic chaos. A systematic study of this relation requires: (a) the identification as well as the quantification of chaos, and (b) the compact representation of particle orbits in a kinetic parameter space. 

In this work, we introduce the SALI (Smaller ALignement Index) as an efficient measure for detecting and quantifying both magnetic and kinetic chaos and discuss its advantages in comparison with other standard measures, such as the mLE (maximum Lyapunov Exponent). Kinetic and magnetic chaos are compared, in terms of Poincar{\'e} surfaces of section for indicative cases of low- and high- energy particles to justify the need for a systematic detailed comparison which is performed in the three-dimensional kinetic space of the constants of motion. Each point of this space uniquely represents an unperturbed GC orbit and classifies it in terms of being trapped or passing, and confined or lost. Under the presence of perturbations, a specific SALI value is assigned to each point, providing detailed information regarding the specific kinetic characteristics of the orbits that actually become chaotic due to perturbations, as well as their position with respect to the wall of the torus. These diagrams provide a detailed phase space resolution of kinetic chaos, enabling the physical intuition on the role of specific sets of perturbative modes in terms of particle, energy and momentum transport, and can be employed as a valuable tool for understanding the interplay between internal instabilities and intentionally induced external perturbations, and investigate synergetic effects and mode-engineering strategies for mitigating high-energy particle losses. Moreover, they are of particular importance in cases of multiple time-dependent perturbations where Poincar{\'e} surfaces of section are difficult to be defined and used for the visualization of the phase space topology. Overall, the quantification of kinetic chaos contains distilled information of the complex phase space dynamics that can be further used in integrated tokamak transport simulation codes, as well as for comparison with fast ion and other diagnostics.              

\section{\label{sec: Acknowledgments} Acknowledgments}
We thank the three anonymous referees, whose comments allowed us to improve the clarity of our work. This work has been carried out within the framework of the EUROfusion Consortium, funded by the European Union via the Euratom Research and Training Programme (Grant Agreement No 101052200 — EUROfusion). However, the views and opinions expressed are those of the author(s) only and do not necessarily reflect those of the European Union or the European Commission. Neither the European Union nor the European Commission can be held responsible for them. The work has also been partially supported by the National Fusion Programme of the Hellenic Republic – General
Secretariat for Research and Innovation, and the Erasmus+ program. H.~T.~M.~ also acknowledges support by a PhD Fellowship from the Science Faculty of the University of Cape Town and partial funding by the UCT Incoming International Student Award as well as the Ethiopian Ministry of Education and Woldia University. H.~T.~M.~and Ch.~S.~thank the Centre for High-Performance Computing (CHPC) of South Africa, for providing the computational resources needed for obtaining the numerical results of this work.

\nocite{*}
\providecommand{\noopsort}[1]{}\providecommand{\singleletter}[1]{#1}%
%


\end{document}